\documentclass[useAMS,usenatbib]{mn2e}
\usepackage[dvips]{graphicx}
\usepackage{txfonts}
\usepackage{amssymb}
\usepackage{indentfirst}
\begin{document}

\title[Unresolved X-ray emission and progenitors of Classical Novae]{Unresolved X-ray emission in M31 and constraints on progenitors of Classical Novae}

\author[\'A. Bogd\'an \& M. Gilfanov]{\'A. Bogd\'an$^{1}$\thanks{E-mail:
bogdan@mpa-garching.mpg.de (\'AB); gilfanov@mpa-garching.mpg.de (MG)} 
and
M. Gilfanov$^{1,2}$\footnotemark[1]\\
$^{1}$Max-Planck-Institut f\"ur Astrophysik, Karl-Schwarzschild-Str.1,
85741 Garching bei M\"unchen, Germany\\
$^{2}$Space Research Institute, Russian Academy of Sciences, Profsoyuznaya
84/32, 117997 Moscow, Russia}

\date{}

\maketitle

\begin{abstract}
We investigate unresolved X-ray emission from M31 based on an extensive set of archival \textit{XMM-Newton} and \textit{Chandra} data. We show that extended emission, found previously in the bulge and thought to be associated with a large number of faint compact sources, extends to the disk of the galaxy with similar X-ray to K-band luminosity ratio. We also detect excess X-ray emission associated with the 10-kpc star-forming ring. The  $ \mathrm{L_X/SFR}$  ratio in the $0.5-2$ keV band ranges from zero to 
$\approx 1.8 \cdot 10^{38} \ \mathrm{(erg \ s^{-1})/(M_{\odot}/yr)} $, excluding the regions near the minor axis of the galaxy where it is $\sim 1.5-2$ times higher. The latter is likely associated with warm ionized gas of the galactic wind rather than with the star-forming ring itself.

Based on this data, we constrain the nature of Classical Nova (CN) progenitors. We use the fact that hydrogen-rich material,  required to trigger the explosion, accumulates on the white dwarf surface via accretion. Depending on the type of the system, the energy of accretion may be radiated at X-ray energies, thus contributing to the unresolved X-ray  emission. Based on  the CN rate in the bulge of M31 and its X-ray surface brightness, we show that no more than $ \sim 10 $ per cent of CNe can be produced in magnetic cataclysmic variables, the upper limit being $ \sim 3 $ per cent for parameters typical for CN progenitors.  In dwarf novae, $\gtrsim 90-95  $ per cent of the material must be accreted during outbursts, when the emission spectrum is soft, and only a small fraction  in quiescent periods, characterized by rather hard spectra.
\end{abstract}


\section{Introduction}
Similarly to other normal galaxies, X-ray emission from the bulge of the Andromeda galaxy is dominated by accreting compact sources \citep[e.g.][]{voss}. In addition, there is relatively bright extended emission which nature was explored in \citet{li,bogdan} (hereafter Paper I) based on extensive set of \textit{Chandra} observations. In Paper I we revealed the presence of warm ionized ISM  in the bulge which most likely forms a galactic-scale outflow and showed that bulk of unresolved emission is associated with old stellar population, similar to the Galactic ridge X-ray emission in the Milky Way \citep{revnivtsev2,sazonov}. Although these studies led to a much better understanding of X-ray emission from the bulge of M31, the disk of the galaxy could not be investigated in similar detail due to insufficient \textit{Chandra} data  coverage of the galaxy. The \textit{XMM-Newton} data, available at the time, also did not provide adequate coverage of the galaxy either, with  only the northern part of the disk observed with relatively short exposures \citep{trudolyubov}. 

Over the last several years \textit{XMM-Newton} completed a survey of M31, which data has become publicly available now.  The numerous pointings with  total exposure time of  $\sim 1.5 $ Ms give a good coverage of the entire galaxy. With these data it has become possible  to study  X-ray emission from the disk of the galaxy. In the present paper we concentrate on the unresolved emission component  and compare its characteristics in the disk and bulge. We also study the 10-kpc star-forming ring to support our earlier claim that   excess unresolved emission is associated with star-forming  regions in the galactic disk (Paper I).

\begin{figure*}
\hbox{
\includegraphics[width=8.5cm]{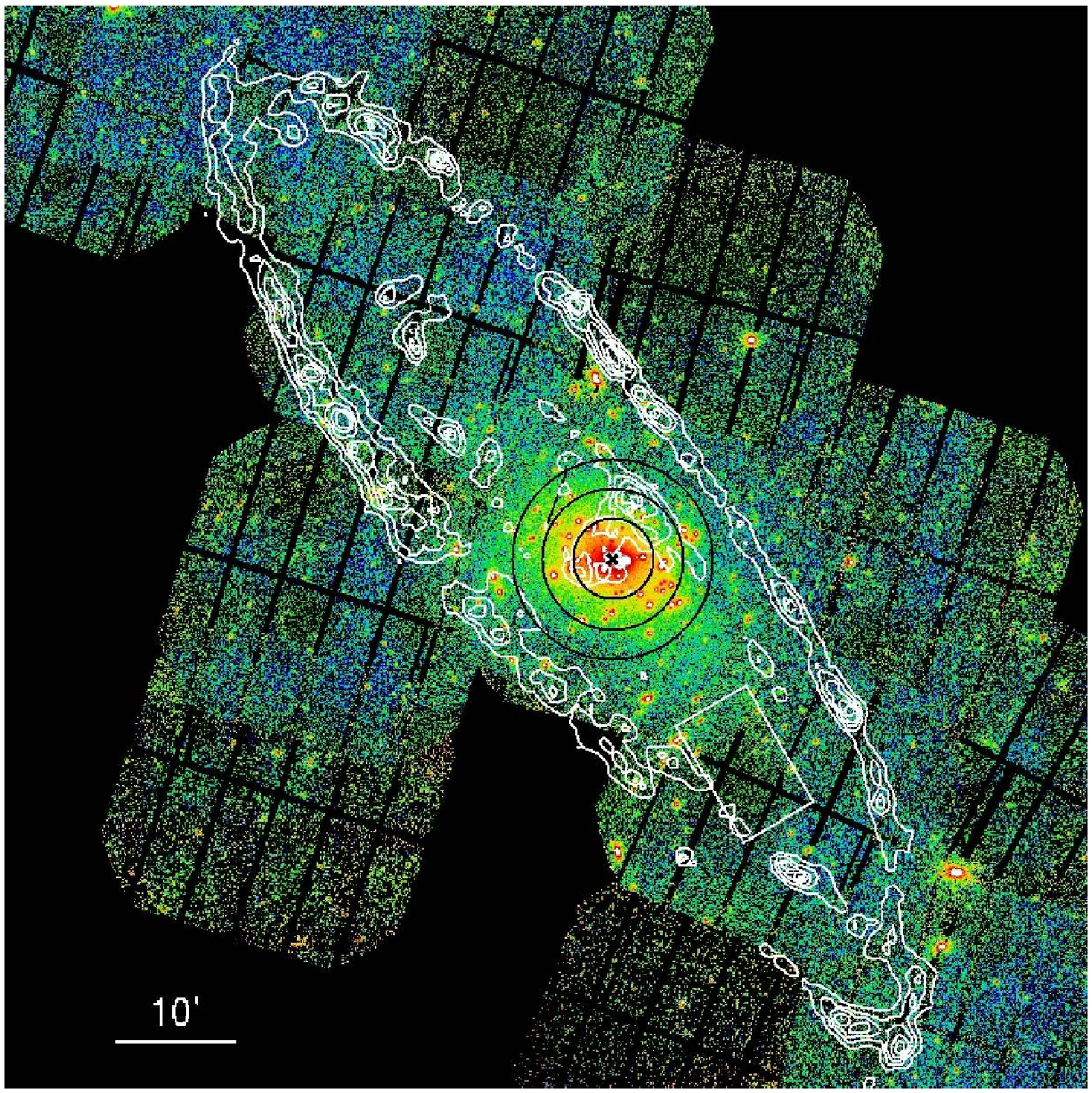}
\hspace{0.45cm}
\includegraphics[width=8.5cm]{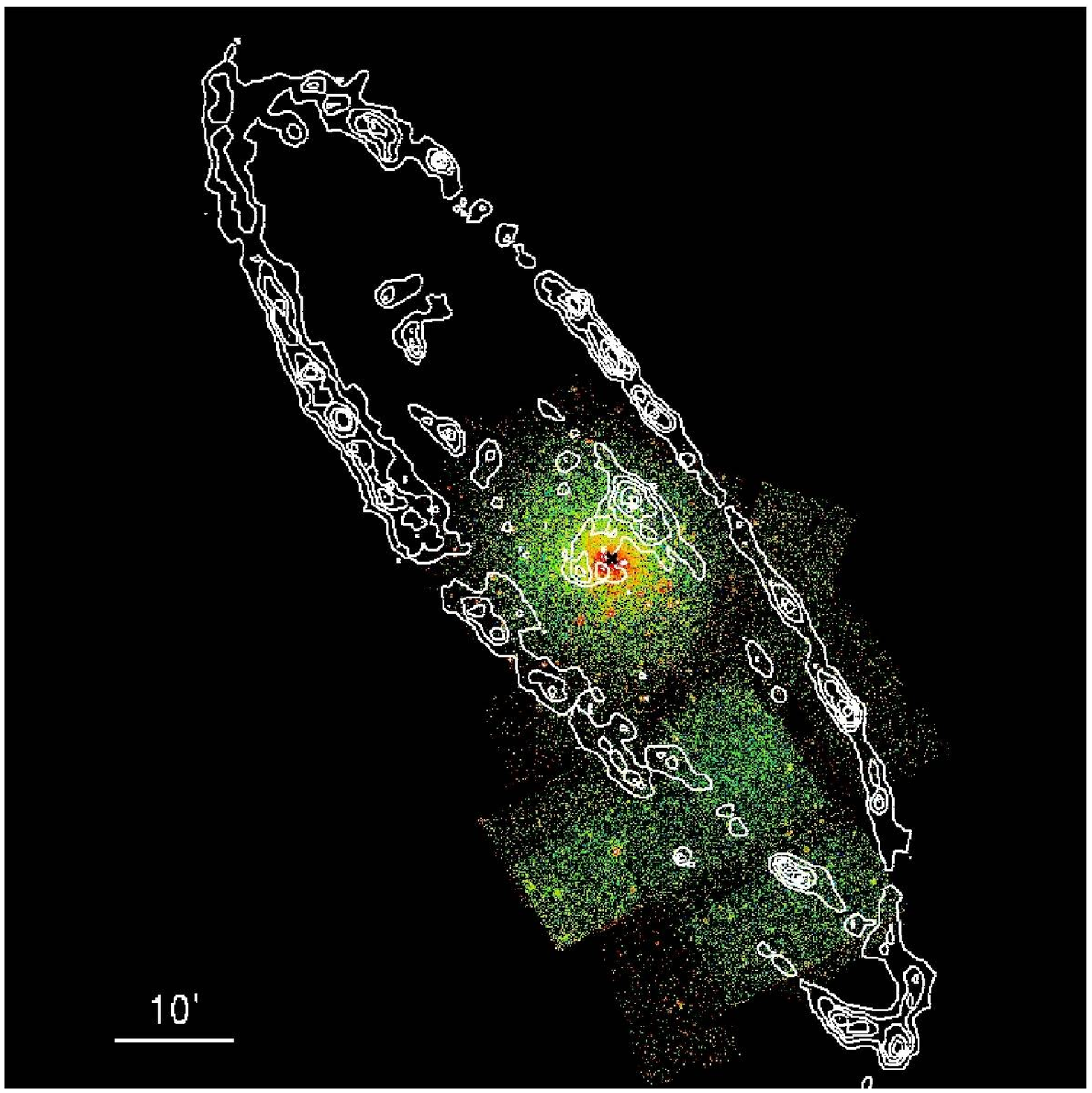}
}
\caption{Combined image of \textit{XMM-Newton} (left panel) and \textit{Chandra} (right) observations in the $ 0.5-2 \ \mathrm{keV} $ energy band. The instrumental background components are subtracted, and the telescope vignetting correction is applied. White contours show the location of the 10-kpc star-forming ring, traced by the $ 160 \ \mathrm{\mu m} $ \textit{Spitzer} image. Overplotted are the  regions used for spectral analysis in Section \ref{sec:spec_anal} and to compute the $ L_X/L_K $ ratios in Section \ref{sec:xtok}. The center of M31 is marked with the cross. North is up and east is left.}
\label{fig:bigm31}
\end{figure*}

Classical Nova explosions are caused by thermonuclear runaway on the surface of a white dwarf (WD) in a binary system \citep{starrfield}. In order for the nuclear runaway to start, a certain amount of hydrogen rich material, $\Delta M\sim 10^{-5} \ \mathrm{M_\odot}$, needs to be accumulated on the WD surface \citep{fujimoto}. This material is supplied by the donor star and is accreted onto the WD. Obviously, there is a direct relation between the frequency of CNe and collective accretion rate in their progenitors.
The accretion energy is released in the form of electromagnetic radiation which spectrum depends on the type of the progenitor system. In certain types of accreting WDs it peaks in the X-ray band, for example in magnetic systems -- polars and intermediate polars. Emission from these systems contributes to the unresolved emission in the galaxy, therefore their number and  contribution to the CN rate can be constrained using high resolution X-ray data. This is the subject of the second part of the paper. Note, that a similar line of arguments can be used to constrain the nature of progenitors of Type Ia Supernovae \citep{nature}.

In the following, the distance to M31 is assumed to be  $ 780 $ kpc \citep{stanek,macri} and the  Galactic hydrogen column density  is $ 6.7 \times 10^{20} \ \mathrm{cm^{-2}} $ \citep{dickey}. 

The paper is structured as follows. In Section 2 we describe the analyzed X-ray and near-infrared data and the main steps of its reduction. In Section 3 results of the analysis of unresolved X-ray emission are presented. In Section 4 we derive  constraints on the nature of CN progenitors. Our results are summarized in Section 5.

\section{Data reduction}
\subsection{\textit{XMM-Newton}}
\label{sec:xmm}
We analyzed $ 24 $ observations from \textit{XMM-Newton} survey of M31, listed in Table \ref{tab:xmm}. The data was taken between 2000 June 25 and 2007 July 25 and covers nearly the entire disk of M31 and its bulge. The approximate coverage of the galaxy by \textit{XMM-Newton} data is illustrated in the left panel of Fig. \ref{fig:bigm31}. We analyzed the data of the European Photon Imaging Camera (EPIC) instruments \citep{struder,turner}, for its reduction we used Science Analysis System (\begin{small}SAS\end{small}) version 7.1. 

\begin{table}
\caption{The list of \textit{XMM-Newton} observations used for the analysis.}
\centering
\begin{tabular}{c|c|c|c}
\hline
Obs-ID & $ T_{\mathrm{obs}} $ (ks) &  $ T_{\mathrm{filt}} $ (ks) &  Date  \\
\hline
$ 0109270101 $  & $ 62.5 $ & $ 16.0 $ & 2001 Jun 29  \\
$ 0109270301 $  & $ 54.2 $ & $ 25.2 $ & 2002 Jan 26  \\
$ 0109270401 $  & $ 91.6 $ & $ 32.2 $ & 2002 Jun 29  \\
$ 0109270701 $  & $ 54.9 $ & $ 53.7 $ & 2002 Jan 05  \\
$ 0112570101 $  & $ 61.1 $ & $ 52.6 $ & 2002 Jan 06  \\   
$ 0112570201 $  & $ 62.8 $ & $ 42.6 $ & 2002 Jan 12  \\   
$ 0112570301 $  & $ 59.9 $ & $ 15.2 $ & 2002 Jan 24  \\   
$ 0112570401 $  & $ 31.0 $ & $ 25.0 $ & 2000 Jun 25  \\   
$ 0202230201 $  & $ 18.3 $ & $ 17.8 $ & 2004 Jul 16  \\   
$ 0202230401 $  & $ 14.6 $ & $ 9.0  $  & 2004 Jul 18  \\
$ 0202230501 $  & $ 21.8 $ & $ 2.0  $  & 2004 Jul 19  \\   
$ 0402560301 $  & $ 66.7 $ & $ 33.9 $ & 2006 Jul 01  \\   
$ 0402560401 $  & $ 57.0 $ & $  5.0 $ & 2006 Jul 08  \\   
$ 0402560501 $  & $ 57.0 $ & $ 24.0 $ & 2006 Jul 20  \\   
$ 0402560701 $  & $ 64.4 $ & $ 14.0 $ & 2006 Jul 23  \\   
$ 0402560801 $  & $ 64.0 $ & $ 41.0 $ & 2006 Dec 25  \\   
$ 0402560901 $  & $ 60.0 $ & $ 33.6 $ & 2006 Dec 26  \\   
$ 0402561001 $  & $ 63.6 $ & $ 39.7 $ & 2006 Dec 30  \\   
$ 0402561101 $  & $ 60.0 $ & $ 32.0 $ & 2007 Jan 01  \\   
$ 0402561301 $  & $ 52.1 $ & $ 25.1 $ & 2007 Jan 03  \\   
$ 0402561401 $  & $ 62.0 $ & $ 40.0 $ & 2007 Jan 04  \\   
$ 0402561501 $  & $ 54.9 $ & $ 36.9 $ & 2007 Jan 05  \\   
$ 0404060201 $  & $ 40.0 $ & $ 14.0 $ & 2006 Jul 03  \\   
$ 0410582001 $  & $ 20.1 $ & $  8.0 $ & 2007 Jul 25  \\   

\hline
\end{tabular}
\label{tab:xmm}
\end{table} 

Main steps of data analysis were performed in the same way as described in Paper I. After applying the double filtering technique \citep{nevalainen} the exposure time decreased to $ T_{\mathrm{filt}} \approx 639 $ ks. The out-of-time events were removed using the Oot event list. The exposure maps were calculated with \begin{small}EEXPMAP\end{small}
command of \begin{small}SAS\end{small}, using a powerlaw model with slope of $ \Gamma =2 $ The observations were re-projected and merged in the coordinate system of Obs-ID 0112570101. 

In the analysis of extended emission, contribution of resolved point sources needs to be removed. For this,  we used the Chandra source list, where available.  In the disk, not covered by Chandra, we ran the \begin{small}SAS\end{small} source detection tool, which gives a complete list of sources above $ \sim 10^{36} \ \mathrm{erg \ s^{-1}} $. The resulting list was used to mask  out compact sources in these regions. As before (Paper I), we use enlarged source regions in order to limit the contribution of remaining counts from point sources to less than $\lesssim 10$ per cent. 

The particle background components were subtracted as described in Paper I, where their origin is also discussed. By removal of the cosmic X-ray background (CXB) we took into account that the resolved fraction of CXB changes as the exposure time varies. In order to  compensate for this effect we assumed that the point source detection sensitivity is proportional to the exposure time, and computed the resolved fraction of CXB for each pixel using ($ \log N - \log S $) distribution from  \citet{moretti}. We took into account various systematic errors in the background subtraction procedure, which include the scatter in the determination of the ``flat'' internal background (caused by the interaction of cosmic rays with the detector material), the uncertainty in the ($ \log N - \log S $) distribution, and the uncertainty in the determination of the solar proton component. The combined effect of these uncertainties is indicated by grey shaded area in the surface brightness profiles analyzed in Section \ref{sec:prof}.  Due to limitations of the background subtraction procedure, data can be analyzed and interpreted  reliably only out  to $ \sim 1 \degr $ central distance along the major axis of M31 in the $ 0.5 - 2 \ \mathrm{keV} $ energy range. We found that  in harder energy band only the bulge of the galaxy can be studied with XMM-Newton data.

\begin{table}
\caption{The list of \textit{Chandra} observations used for the analysis.}
\centering
\begin{tabular}{c|c|c|c|c}
\hline
Obs-ID & $ T_{\mathrm{obs}} $ (ks) &  $ T_{\mathrm{filt}} $ (ks) & Instrument & Date  \\
\hline
$ 303 $  & $ 12.0 $ &$ 8.2    $  & ACIS-I & 1999 Oct 13  \\
$ 305 $  & $ 4.2 $  & $ 4.0   $ & ACIS-I & 1999 Dec 11  \\   
$ 306 $  & $ 4.2 $  & $ 4.1   $ & ACIS-I & 1999 Dec 27  \\   
$ 307 $  & $ 4.2 $  & $ 3.1   $ & ACIS-I & 2000 Jan 29  \\   
$ 308 $  & $ 4.1 $  & $ 3.7   $ & ACIS-I & 2000 Feb 16  \\
$ 311 $  & $ 5.0 $  & $ 3.9   $ & ACIS-I & 2000 Jul 29  \\   
$ 312 $  & $ 4.7 $  & $ 3.8   $ & ACIS-I & 2000 Aug 27  \\   
$ 313 $  & $ 6.1 $  & $ 2.6   $ & ACIS-S & 2000 Sep 21  \\   
$ 314 $  & $ 5.2 $  & $ 5.0   $ & ACIS-S & 2000 Oct 21  \\   
$ 1575 $ & $ 38.2 $ & $ 38.2  $ & ACIS-S & 2001 Oct 05  \\
$ 1577 $ & $ 5.0 $  & $ 4.9   $ & ACIS-I & 2001 Aug 31  \\
$ 1580 $ & $ 5.1 $  & $ 4.8   $ & ACIS-S & 2000 Nov 17  \\
$ 1583 $ & $ 5.0 $  & $ 4.1   $ & ACIS-I & 2001 Jun 10  \\
$ 1585 $ & $ 5.0 $  & $ 4.1   $ & ACIS-I & 2001 Nov 19  \\   
$ 2049 $ & $ 14.8 $ & $ 11.6  $  & ACIS-S & 2000 Nov 05  \\   
$ 2050 $ & $ 13.2 $ & $ 10.9  $  & ACIS-S & 2001 Mar 08  \\   
$ 2051 $ & $ 13.8 $ & $ 12.8  $  & ACIS-S & 2001 Jul 03  \\   
$ 2895 $ & $ 4.9 $  & $ 3.2   $ & ACIS-I & 2001 Dec 07  \\   
$ 2896 $ & $ 5.0 $  & $ 3.7   $ & ACIS-I & 2002 Feb 06  \\
$ 2897 $ & $ 5.0 $  & $ 4.1   $ & ACIS-I & 2002 Jan 08  \\
$ 2898 $ & $ 5.0 $  & $ 3.2   $ & ACIS-I & 2002 Jun 02  \\
$ 4360 $ & $ 5.0 $  & $ 3.4   $ & ACIS-I & 2002 Aug 11  \\   
$ 4536 $ & $ 54.9 $ & $ 30.2  $  & ACIS-S & 2005 Mar 07  \\   
$ 4678 $ & $ 4.9 $  & $ 2.7   $ & ACIS-I & 2003 Nov 09  \\    
$ 4679 $ & $ 4.8 $  & $ 2.7   $ & ACIS-I & 2003 Nov 26  \\
$ 4680 $ & $ 5.2 $  & $ 3.2   $ & ACIS-I & 2003 Dec 27  \\
$ 4681 $ & $ 5.1 $  & $ 3.3   $ & ACIS-I & 2004 Jan 31  \\   
$ 4682 $ & $ 4.9 $  & $ 1.2   $ & ACIS-I & 2004 May 23  \\
$ 7064 $ & $ 29.1 $ & $ 23.2  $  & ACIS-I & 2006 Dec 04  \\
$ 7068 $ & $  9.6 $ & $  7.7  $  & ACIS-I & 2007 Jun 02  \\
\hline                                       
\end{tabular}                                
\label{tab:chand}
\end{table}

\subsection{\textit{Chandra}}

We combined an extensive set of \textit{Chandra} data taken between 1999 October 13 and 2007 June 2, listed in Table \ref{tab:chand}. The data was processed with \begin{small}CIAO\end{small}\,\footnote[1]{http://cxc.harvard.edu/ciao/} software package tools (\begin{small}CIAO\end{small} version 4.0; \begin{small}CALDB\end{small} version 3.4.5). The $ 29 $ observations allowed us to study the bulge and the southern disk of M31, the exact data coverage is shown in the right panel of Fig. \ref{fig:bigm31}. In case of ACIS-S observations we used the data from S1, S2, S3, I2, I3 CCDs, except for Obs-ID 1575 where we extracted data only from the S3 chip. The peak value of the exposure time reaches $ T_{filt} \approx 144 \ \mathrm{ks} $ in the center of the galaxy.

The main steps of the data analysis are similar to those outlined in Paper I. After excluding the flare-contaminated time intervals the total exposure time decreased by $\sim 25  $ per cent. The instrumental background was subtracted following \citet{hickox} and using the stowed data set\footnote[2]{http://cxc.harvard.edu/contrib/maxim/stowed/}. By subtraction of CXB we followed the same procedure as described for \textit{XMM-Newton}, to correct for the varying fraction of resolved CXB sources.

\begin{figure}
\resizebox{\hsize}{!}{\includegraphics{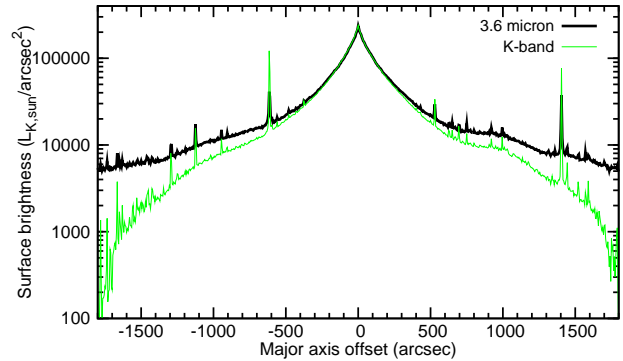}}
\caption{Near-infrared light distribution along the major axis of M31 based on $ 3.6 \ \mathrm{\mu m} $ data of \textit{Spitzer Space Telescope} (black thick line) and the \textit{2MASS} K-band (thin green line) data. The normalization of \textit{Spitzer} profile was adjusted to match the K-band light in the center of M31. The x-coordinate increases from south-west to north-east.} 
\label{fig:2massspitz}
\end{figure}

\subsection{Near-infrared data}
\label{sec:nir}
In order to compare the unresolved X-ray emission with the stellar mass distribution, a stellar mass tracer is needed. The K-band image from the \textit{Two-Micron All Sky Survey} (\textit{2MASS}) Large Galaxy Atlas (LGA) \citep{jarrett} image is commonly used for this purpose. Alternatively, one could use data of the Infrared Array Camera (IRAC) on the \textit{Spitzer Space Telescope} (\textit{SST}), which provides images at near-infrared wavelengths, among others at $ 3.6 \ \mathrm{\mu m} $. 

\begin{figure*}
\hbox{
\includegraphics[width=8.75cm]{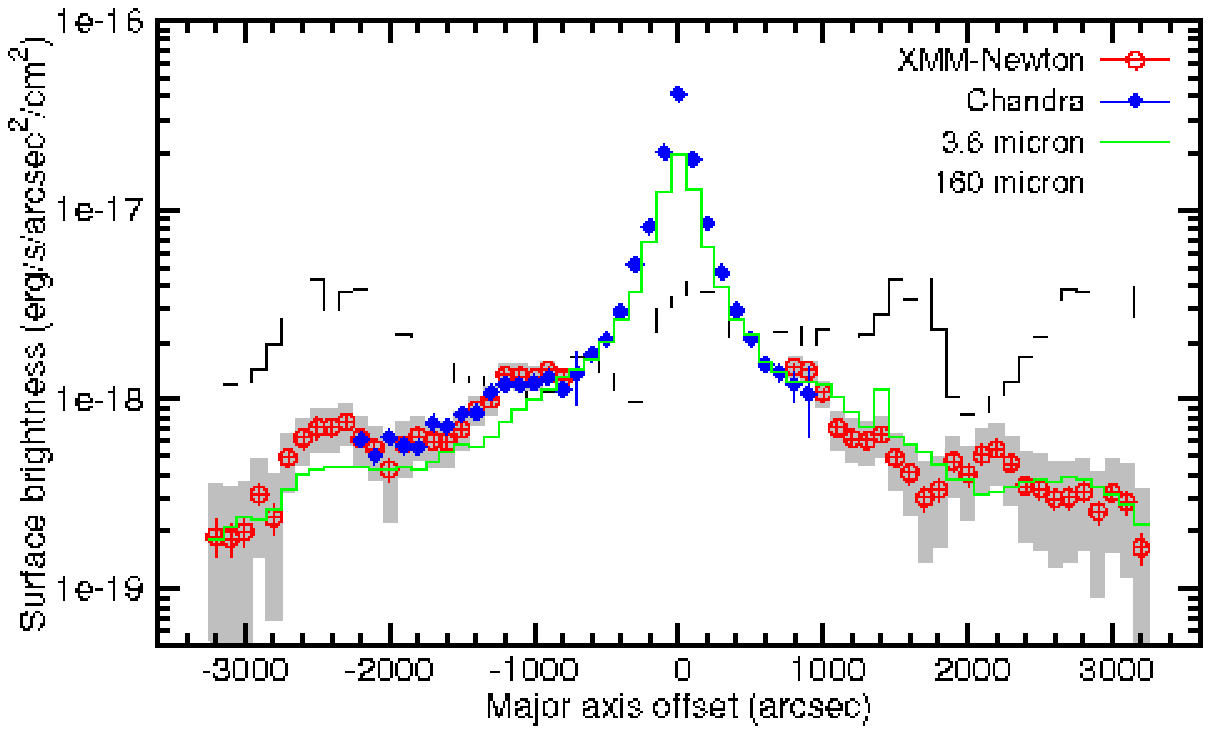}
\hspace{0.25cm}
\includegraphics[width=8.75cm]{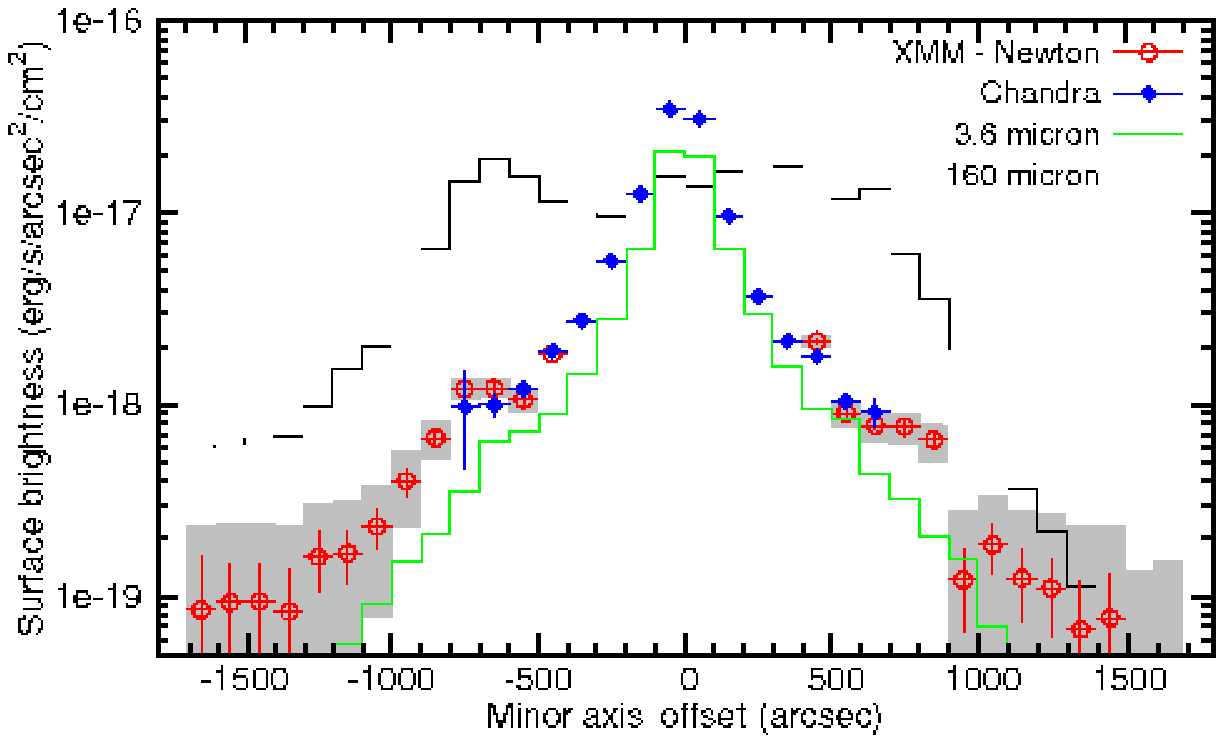}
}
\vspace{-0.3cm}
\caption{X-ray surface brightness distribution along the major  (left panel) and minor axes (right) in the $ 0.5 - 2 \ \mathrm{keV} $ energy band. The filled (blue in the color version) symbols show the \textit{Chandra} data, open (red) symbols represent the \textit{XMM-Newton} data, the solid (green) histogram is the surface brightness of $ 3.6 \ \mathrm{\mu m} $ \textit{Spitzer} data, the thin solid (black) line is  $ 160 \ \mathrm{\mu m} $ \textit{Spitzer} data. The shaded area shows  the systematic uncertainty in the background subtraction for the \textit{XMM-Newton} data. The normalization of  near- and far-infrared profiles are the same on both panels. The x-coordinate increases from south-west to north-east for the major axis and from south-east to north-west for the minor axis profile.} 
\label{fig:bigprof}
\end{figure*}

\begin{figure*}
\hbox{
\includegraphics[width=8.75cm]{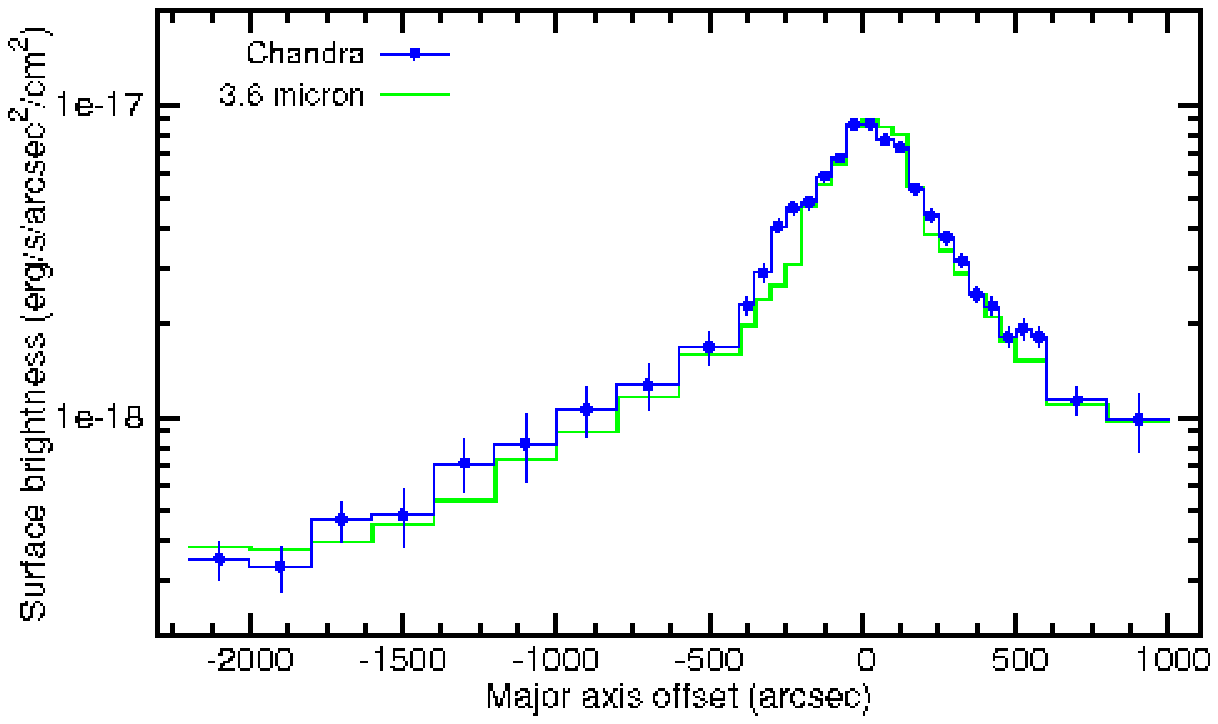}
\hspace{0.25cm}
\includegraphics[width=8.75cm]{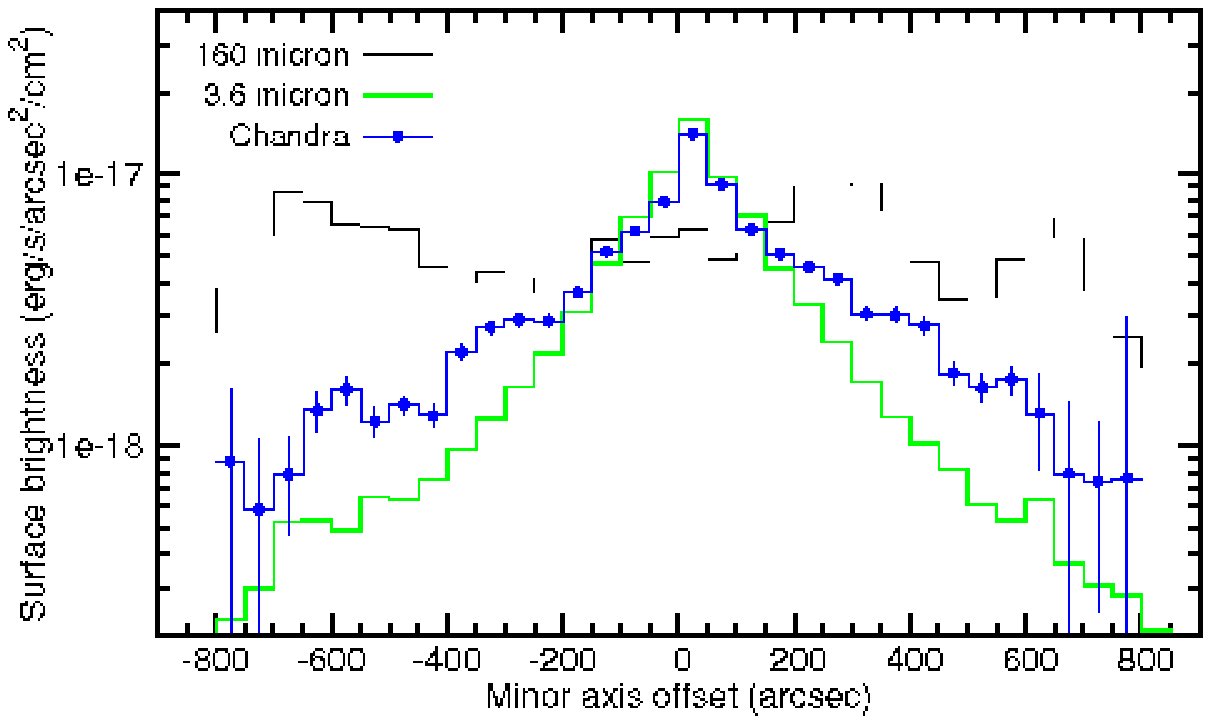}
}
\vspace{-0.3cm}
\caption{Same as Fig. \ref{fig:bigprof} but in the $ 2 - 7 \ \mathrm{keV} $ energy band, based on \textit{Chandra} data. The normalization of the near-infrared profiles is the same on both panels.} 
\label{fig:bigprofhard}
\end{figure*}

Due to the large angular extent of M31 the background on the \textit{2MASS} K-band image provided by LGA is somewhat oversubtracted (T. Jarrett, private communication), therefore the disk of M31 appears to be too faint. In Fig. \ref{fig:2massspitz} we compare the K-band surface brightness distribution with the \textit{Spitzer} $ 3.6 \ \mathrm{\mu m} $ data. The profiles were extracted along the major axis of M31, with bin width of $ 5 \arcsec $ and the bins were averaged on $ 500 \arcsec $ in the transverse direction. The background level of  \textit{Spitzer} image was estimated using nearby fields to M31.  There is a good agreement in the central $  \sim 300 \arcsec $, where the background subtraction is nearly negligible, but at large offsets an increasing deviation appears.  Therefore in the present study we use the $ 3.6 \ \mathrm{\mu m} $ image of \textit{Spitzer} to trace the stellar mass. In order to facilitate the comparison of results with other studies we converted \textit{Spitzer} near-infrared luminosities to K-band values using the central regions of the galaxy. The obtained conversion factor between pixel values is $ C_{K}/C_{3.6 \ \mathrm{\mu m} } \approx 10.4$. 

In Paper I we used 2MASS image to compute K-band luminosities and stellar masses of different parts of the galaxy.  As the analysis of the Paper I did not extend beyond $\approx 1000\arcsec$ from the center, the underestimated disk brightness  on the 2MASS image at large offset angles has not affected our results in any significant or qualitative way. It resulted, however, in small quantitative  difference in X/K ratios between this paper and Paper I.

\section{Unresolved X-ray emission in M31}
\subsection{Surface brightness distribution}
\label{sec:prof}
We study the brightness distribution of unresolved X-ray emission in M31 in the $ 0.5 - 2 \ \mathrm{keV} $ and in the $ 2 - 7 \ \mathrm{keV} $ energy range (Fig.\ref{fig:bigprof},  \ref{fig:bigprofhard}). In all cases profiles were averaged  over $ 500 \arcsec $ in the transverse direction. The contribution of resolved point sources is removed. The profiles are corrected for vignetting, furthermore all instrumental and sky background components are subtracted. We found good agreement at all studied distances between \textit{Chandra} and \textit{XMM-Newton} data. In the inner bulge region, which is most crowded with point sources we consider only \textit{Chandra} data.  X-ray light distributions are compared with the $ 3.6 \ \mathrm{\mu m} $ \textit{Spitzer} data. 

The left panel of Fig. \ref{fig:bigprof} presents the surface brightness distribution along the major axis of M31 in the $0.5-2 $ keV band. The shaded area shows the background subtraction uncertainties (see Section \ref{sec:xmm}). The profile confirms the presence of the additional soft emission in the central bulge, which was shown to originate from hot ionized gas (see Paper I). Outside few central bins X-ray flux follows the near infrared profile -- unresolved emission associated with the bulge of the galaxy continues into the disk with approximately  the same X-ray to K-band ratio. 
Although overall correlation between X-ray and near infrared brightness is quite good, there are several deviations, of which the most prominent is the excess X-ray emission at the central distance of $-2500 \arcsec$. This  excess approximately coincides with the broad hump on the $160 \ \mu$m profile corresponding to the southern end of the 10-kpc star-forming ring (Fig.\ref{fig:bigm31}).  Similar excess emission ("shoulders" at $\sim \pm 700\arcsec$) coinciding with the peaks on the $160 \ \mu$m profiles is present in the distribution along the  minor axis. (Note that overall normalization of the X-ray flux is larger on the minor axis profile due to contribution of the gas emission.)
A possible association of excess X-ray emission with the star-forming activity in the 10-kpc star-forming ring is further discussed in Section \ref{sec:xsfr}.

In Fig. \ref{fig:bigprofhard} we show the brightness distribution in the $ 2-7 $ keV energy band based on \textit{Chandra} data. No contribution of the hot gas emission with $kT\sim 300-400$ eV is expected in this energy range. Accordingly,  there is a good agreement between the X-ray and near-infrared distribution along the major axis (Fig. \ref{fig:bigprofhard}).  However, along the minor axis  the unresolved X-ray emission traces the stellar light only in the inner region. Outside  $\sim 200-300 \arcsec $ the X-ray brightness becomes systematically larger than the normalized $ 3.6 \ \mathrm{\mu m}$ distribution. The origin of this enhancement is not clear. It may be associated with the star-formation activity in the disk and 10-kpc star-forming ring, as suggested in Paper I, or may be related to the galactic scale wind.

\subsection{Spectra}
\label{sec:spec_anal}

In order to extract spectra of the bulge and  the disk of M31 we used only \textit{Chandra} data. The background was subtracted using the ACIS ``blank-sky'' files as described in Paper I. We extracted the spectrum of the inner bulge, outer bulge, and disk of M31. The inner bulge region is represented by a circle with $ 200  $ arcsec radius, the outer bulge spectrum is extracted from a circular annuli with radii of $ 350-500  $ arcsec, and the disk spectrum is computed from a rectangular region at the southern part of the disk. The applied regions are also depicted in Fig \ref{fig:bigm31}. As a comparison we also added M32, the extraction region is same as described in Paper I. In order to compare the spectra we normalized them to the same level of near-infrared luminosity. 

\begin{figure}
\resizebox{\hsize}{!}{\includegraphics{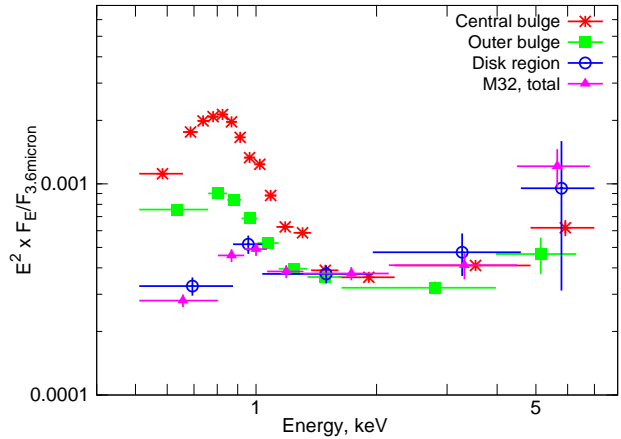}}
\vspace{-0.5cm}
\caption{X-ray spectra of different regions in M31 and in M32: stars (red) show the spectrum of the central $ 200 \arcsec $ region; filled boxes (green) are the spectrum of the  $ 350-500 \arcsec $ annulus; open circles (blue) are the spectrum of the disk; filled triangles (purple) show the spectrum of M32. All spectra were extracted using \textit{Chandra} data. The spectra are normalized to the same level of near-infrared brightness.} 
\label{fig:spec1}
\end{figure}

Fig. \ref{fig:spec1} reveals that all spectra are consistent above $ \sim 1.5 $ keV. Below this energy the inner bulge region is strikingly different, it has a factor of $\sim 4 $ times stronger soft component than all other spectra. Also the outer bulge has a weak excess below $ \sim 1 $ keV compared to the disk and M32 spectra, which show very similar spectral characteristics at all energies. The remarkable soft component is the consequence of the hot ionized gas, located in the bulge of M31 (see Paper I). The somewhat increased soft emission in the outer bulge is presumably due to the contribution from gas emission, lacking completely in the disk of the galaxy.  A relatively weak soft component is also present in the spectra of the disk region and M32, which presumably originates from the population of unresolved sources. Indeed, the different nature of the soft components in the inner bulge and in the disk of M31 is also supported by difference in their best-fit temperatures. Using a simple two component spectral model, consisting of an optically-thin thermal plasma emission spectrum and a power-law model (MEKAL in XSPEC), we find a best-fit temperature of $ kT =0.36 \pm 0.01 $ keV for the inner bulge, whereas we obtain $ kT = 0.62 \pm 0.10 $ keV in the disk and $ kT = 0.54 \pm 0.15 $ keV in M32.

\label{sec:spec}

\begin{table*}
\caption{X-ray to K-band luminosity ratios for different regions of M31 and for M32.}
\renewcommand{\arraystretch}{1.2}
\centering
\begin{tabular}{c c c c c c}
\hline
Region &  $ L_K  $ & $ L_{0.5-2keV} $  &  $ L_{2-10keV} $ & $ L_{0.5-2keV}/L_K  $ & $ L_{2-10keV}/L_K  $ \\
 & ($ \mathrm{L_{K,\odot}} $) & ($  \mathrm{erg \ s^{-1}}$) & ($  \mathrm{erg \ s^{-1}}$) & ($ \mathrm{erg \ s^{-1} \ L_{K,\odot}^{-1}} $) &  ($ \mathrm{erg \ s^{-1} \ L_{K,\odot}^{-1}} $) \\
\hline
Inner bulge  & $ 1.9 \cdot 10^{10} $  & $ 1.7 \cdot 10^{38}  $  & $  8.0 \cdot 10^{37}  $ & $  (8.9 \pm 0.1) \cdot 10^{27}  $ & $  (4.2 \pm 0.1) \cdot 10^{27}  $ \\
Outer bulge  & $ 8.2 \cdot 10^{9}  $  & $ 4.1  \cdot 10^{37} $  & $  2.9 \cdot 10^{37}  $ & $  (5.0 \pm 0.1) \cdot 10^{27}  $ & $  (3.4 \pm 0.1) \cdot 10^{27}  $  \\ 
Disk         & $ 2.1 \cdot 10^{9}  $  & $ 7.6  \cdot 10^{36} $  & $  9.4 \cdot 10^{36}  $ & $  (3.6 \pm 0.1) \cdot 10^{27}  $ & $  (4.5 \pm 0.2) \cdot 10^{27}  $  \\   
M32          & $ 8.5 \cdot 10^8    $  & $ 3.0 \cdot 10^{36} $   & $  3.4 \cdot 10^{36} $  & $  (3.5 \pm 0.1) \cdot 10^{36}  $ & $  (4.0 \pm 0.2) \cdot 10^{36}  $  \\
\hline 
\end{tabular}
\label{tab:xtokratio}
\end{table*}

\subsection{$L_X/L_K$ ratios}
\label{sec:xtok}
The X-ray to K-band luminosity ratios ($L_X/L_K$) were computed for the same regions as  used for spectral analysis. As before, we use  \textit{Chandra} data only. The $L_X/L_K$ ratios are obtained in the $ 0.5-2 $ keV and in the $ 2-10 $ keV energy range to facilitate comparison with previous studies. The X-ray luminosities in the $ 0.5-2 $ keV energy range were computed using the best fit spectral models. For the outer bulge and disk we assumed a "power law + MEKAL" model, whereas for the inner bulge a second MEKAL component was added to obtain a better fit.  In the $ 2-10 $ keV band the X-ray luminosities were computed from the count rate using the count-to-erg conversion factor for a power law spectrum with a photon index of $\Gamma=2$.  The conversion factor depends weakly on the assumed slope, changing by $\lesssim 10$ per cent when $\Gamma$ varied by $\pm0.4$. The luminosities and their ratios are  listed in Table \ref{tab:xtokratio},  errors correspond to statistical uncertainties in the X-ray count rates. 

The $L_X/L_K$ ratios for the bulge regions are consistent with  those given in Paper I, albeit  somewhat smaller, due to the increased K-band luminosity for some of the regions (Section \ref{sec:nir}). In agreement with  spectra, the $L_X/L_K$ ratio in the soft band is highest in the inner bulge, and  smallest in the disk of the galaxy, where it is consistent with M32 value (Paper I). In the outer bulge the obtained $L_X/L_K$ ratio is in-between due to residual contribution from gas emission. In the hard band all $L_X/L_K$ ratios are similar to each other, in the range of $ (3.4-4.5) \cdot 10^{27} \  \mathrm{erg \ s^{-1} \ L_{K,\odot}^{-1}} $. There is a statistically significant scatter in their values, which origin is not clear. It can not be explained by the varying residual contribution of LMXBs due to different point source detection sensitivity in different region. Indeed, the latter varies from $ \sim 2 \cdot 10^{35} \ \mathrm{erg \ s^{-1}}  $ in the outer bulge to $ \sim  6 \cdot 10^{35} \ \mathrm{erg \ s^{-1}} $ in the disk. If we use the luminosity function of LMXBs of \citet{gilfanov} and assume a powerlaw spectrum with slope of $ \Gamma =1.56 $ as average LMXB spectrum \citep{irwin}, we find that LMXBs in the luminosity range of $(2-6) \cdot 10^{35} \ \mathrm{erg \ s^{-1}}$ contribute $L_X/L_K \sim 3 \cdot 10^{26} \  \mathrm{erg \ s^{-1} \ L_{K,\odot}^{-1}} $, which may account only for $\sim 1/4$ of the scatter.
Further contribution to the observed difference in $ L_X/L_K  $ ratios may be made by the difference in star-formation history of different regions. This may be an interesting topic on its own, but it is beyond the scope of this paper.

\subsection{Emission from the 10-kpc star-forming ring}
\label{sec:xsfr}
Surface brightness profiles suggest that there may be additional emission component associated with spiral arms and the 10-kpc star-forming ring. In order to study this further, we consider X-ray emission along the 10-kpc star-forming ring of the galaxy and investigate the behavior of the  $L_X$/SFR(star formation rate) ratio. We use  \textit{XMM-Newton} data since only they provide good coverage of the star-forming ring with adequate sensitivity. Due to uncertainties in the background subtraction procedure in the hard energy band, we restrict this study to $0.5-2$ keV range. The X-ray luminosity was computed in the same way as described in Section \ref{sec:prof}.  The contribution of unresolved emission associated with old stellar population was removed based on the near-infrared luminosity of studied regions and using the $L_X/L_K$ ratio of the disk of M31. The remaining X-ray emission is $ 25-50 $ per cent of of the original value.  The star-formation rate was determined based from $ 160 \ \mathrm{\mu m} $   image of the galaxy provided by \textit{Spitzer}. The background level for the latter was determined from a combination of nearby blank sky fields. The star-formation rate was computed from the $160 \mu$m flux  using  the infrared spectral fits from \citet{gordon} and calibration of \citet{kennicutt}. This resulted to a conversion coefficient of  SFR $=  9.5 \cdot 10^{-5} \ \mathrm{F_{160 \mu m}/Jy \ M_{\odot} yr^{-1}}$  for the distance of M31. 

\begin{figure}
\resizebox{\hsize}{!}{\includegraphics{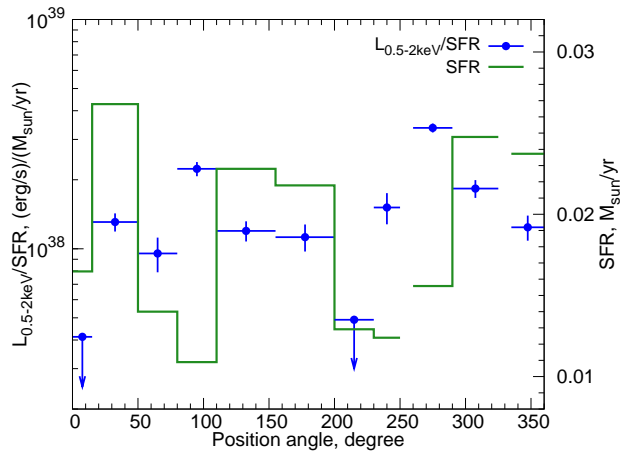}}
\caption{The X/SFR ratio along the 10-kpc star-forming  (filled circles with error bars). The position angle increases clockwise and its zero point is at the northern side of the galaxy along its major axis. The bins centered at $ \approx 90^{\circ} $ and $\approx 270^{\circ} $  approximately correspond to the minor axis of the galaxy.  The solid histogram shows the star formation rate in the same regions (right hand \textit{y}-axis). No X-ray emission above the level defined by   the X/K luminosity ratio of the disk was detected in the bins located at  $ \approx 10^{\circ} $ and $\approx 215^{\circ} $. In order to illustrate the amplitude of possible systematic uncertainties, we show by upper limit signs the level, corresponding to 20\% of the emission from old population.  No X-ray data with sufficient exposure time was available at position angles of $\approx 250^{\circ}$ and  $\approx 330^{\circ}$.}
\label{fig:xtosfr}
\end{figure}

The behavior of  X/SFR ratio along the 10-kpc star-forming ring is shown in Fig. \ref{fig:xtosfr}. It is the  largest in the bins centered at the position angle of $\approx 90^{\circ}$ and $\approx 270^{\circ}$ corresponding to the minor axis of the galaxy. These bins are contaminated by the gas emission as it is obvious from the minor axis soft band profile shown in Fig.\ref{fig:bigprof}.  On the other hand, no excess X-ray emission above the level corresponding to the $L_X/L_K$ ratio for the disk was detected in  the northern (the position angle of $ \sim 10^{\circ} $)  and southern ($ \sim 215^{\circ} $) parts of the ring. In order to indicate the level of possible systematic uncertainties (the statistical errors are much smaller) we show the level corresponding to $20$ per cent of X-ray emission associated with the old stellar population by upper limit signs. Apart from these bins, Fig. \ref{fig:xtosfr} clearly demonstrates presence of the excess unresolved emission approximately correlated with the far-infrared luminosity with X/SFR values in the range $(0.9 - 1.8) \cdot 10^{38} \ \mathrm{(erg \ s^{-1})/(M_{\odot}/yr)}$.
This emission presumably arises from a multitude of unresolved sources associated with star-formation, such as young stellar objects (protostars and pre-main-sequence stars), young stars \citep[e.g.][]{koyama}, low luminosity Be X-ray binaries. Contribution of supernova remnants and hot X-ray emitting gas may also  play a role.
The origin of  observed variations in the X/SFR ratio  is not entirely clear. They   may be  intrinsic, due to the different star-formation history and population age in different parts of the 10-kpc star-forming ring \citep{pavel}, or may be caused by varying column density  which can be as large as few times $ 10^{21} \ \mathrm{cm^{-2}} $ \citep{nieten}. The latter possibility may play particular  role in disappearing of the X-ray emission at the position angle of $ \sim 10^{\circ} $  and $ \sim 215^{\circ} $. Data in the hard band, unaffected by absorption, could discriminate between these two possibilities. To this end, extensive Chandra observations of the 10-kpc star-forming ring would be instrumental.

\section{Progenitors of classical novae in M31}

\subsection{X-ray emission from progenitors of Classical Novae}

\begin{figure}
\resizebox{\hsize}{!}{\includegraphics{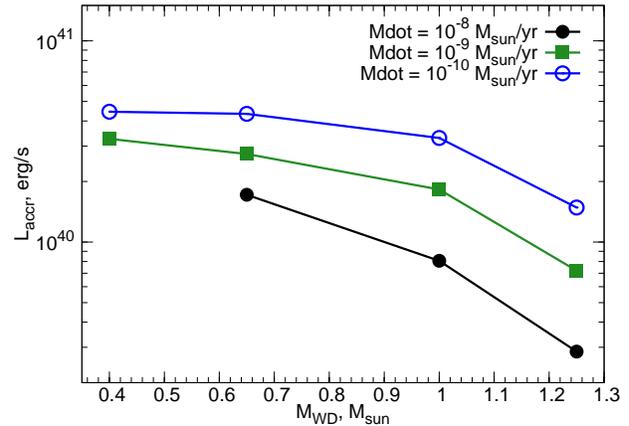}}
\caption{The bolometric luminosity of accreting WDs -- progenitors of Classical Novae in a galaxy  with CN rate of $\nu_{CN}=25$ per year. The curves show dependence of $L_{accr}$ on the WD mass for several values of the mass accretion rate. 
}
\label{fig:cn_lbol}
\end{figure}

Classical Novae (CNe) are nuclear explosions occurring upon accumulation of critical mass of hydrogen-rich material on the surface of an accreting white dwarf. The frequency of these events in a galaxy depends on the rate at which matter accretes onto white dwarfs, hence it can be related to their luminosity.  Indeed, the  accretion energy (bolometric) released between two successive CN outbursts is: 
\begin{equation}
 \Delta  E_{accr} = \frac{G M_{WD} \Delta M}{R_{WD}} \
\end{equation}
where $  \Delta M (M_{WD},\dot{M})  $ is the mass of hydrogen-rich material needed to trigger a CN explosion \citep{yaron}, $ M_{WD} $ and $ R_{WD} $ are the mass and radius of the white dwarf.  For parameters believed to be typical for CN progenitors ($M_{WD} =1 \ \mathrm{M_{\odot}} $ and $ \dot{M} =  10^{-9} \ \mathrm{M_{\odot}/yr}$) $\Delta M\approx 4.7 \cdot 10^{-5} \ \mathrm{M_\odot}$ and  total accretion energy released by an accreting WD between two  CN explosions is $\Delta E_{accr} \sim 2\cdot 10^{46} $ ergs. 
If the frequency of CN events in a galaxy is  $\nu_{CN}$, the total accretion luminosity due to CN progenitors is 
\begin{equation}
L_{accr}= \Delta  E_{accr}\times \nu_{CN} \ .
\label{eq:cn_lbol}
\end{equation}
Dependence of the accretion luminosity on the mass of the WD and accretion rate is shown in Fig.\ref{fig:cn_lbol}.  In computing this curve we used the WD mass-radius relation of  \citet{panei} for a zero temperature  carbon white dwarf and $\Delta M (M_{WD},\dot{M})$ dependence from \citet{yaron}.

The energy of accretion is radiated in the optical, ultraviolet or X-ray bands, depending on the type of the progenitor system. In magnetic systems (polars and intermediate polars) and dwarf novae in quiescence it is emitted predominantly in the X-ray band. Moreover,  their X-ray spectra are relatively hard and their emission is therefore essentially unaffected by the interstellar absorption.  
X-ray radiation from  these objects will contribute to unresolved emission from a galaxy. Therefore, comparing theoretical predictions with the surface brightness of unresolved emission one can constrain their contribution to the observed CN rate in galaxies. Derivation of such upper limits based on M31 data is the goal of this section.

\subsection{Classical Novae, resolved X-ray sources and unresolved  emission in the bulge of  M31}

\label{sec:obsxray}

The CN frequency in the bulge of M31 is  $25\pm4$ \citep{shafter} \citep[see also][]{arp}. 
The number of progenitors required to maintain this rate  is: 
$$ N \sim \frac{\Delta M}{\dot{M}} \nu_{CN}  $$
Typical CN progenitors have accretion rate in the range of $ \sim 10^{-10} -10^{-8} \ \mathrm{M_{\odot}/yr} $ \citep{puebla} and WD masses of   $0.65-1  \ \mathrm{M_{\odot}}$ \citep{ritter}. Correspondingly,  the number of CN progenitors is in the range of $ \sim 2 \cdot 10^3 - 2 \cdot 10^6$, where the smaller number corresponds to the more massive white dwarfs and larger accretion rates. This exceeds significantly the number of bright resolved sources in the bulge of M31, $\sim 300$. The latter  is dominated by  low-mass X-ray binaries -- these being excluded,  the number of potential bright CN progenitors is yet smaller. 
We therefore  conclude that the majority of CN progenitors can not be among bright point sources; rather, they are a part of the unresolved  X-ray emission.

To measure the latter, we define the bulge  as an elliptical region with $ 12 \arcmin $ major axis, with axis ratio of $ 0.47 $ and with position angle of $ 45 \degr $. The total K-band luminosity of the studied region is $ L_{K} =4.2 \cdot 10^{10} \ \mathrm{L_{\odot} } $.  We use \textit{Chandra} data, described in the previous sections, to obtain the unresolved X-ray luminosity in the $ 2-10 $ keV energy range. The advantage of this energy band is that it is not polluted by hot ionized gas (Section \ref{sec:spec_anal}). We find a total luminosity of the unresolved component of $ L_{2-10keV} = (1.6\pm0.1) \cdot 10^{38} \ \mathrm{erg \ s^{-1}} $ in this region. Obviously, this value presents only an upper limit on the luminosity from the population of accreting white dwarfs, as other type of X-ray emitting sources may also contribute. To this end we use results of \citet{sazonov}, who studied the population of faint X-ray sources in the Solar neighborhood and found that  accreting WDs contribute $ \sim 1/3 $ of the total luminosity in the 2--10 keV band. 
As M31 demonstrates similar X/K ratio, we extrapolate this result to the bulge of M31 and estimate the luminosity of CVs of $ L_{CV,2-10keV} = (5.7\pm0.3) \cdot 10^{37} \ \mathrm{erg \ s^{-1}} $.

\begin{figure}
\resizebox{\hsize}{!}{\includegraphics{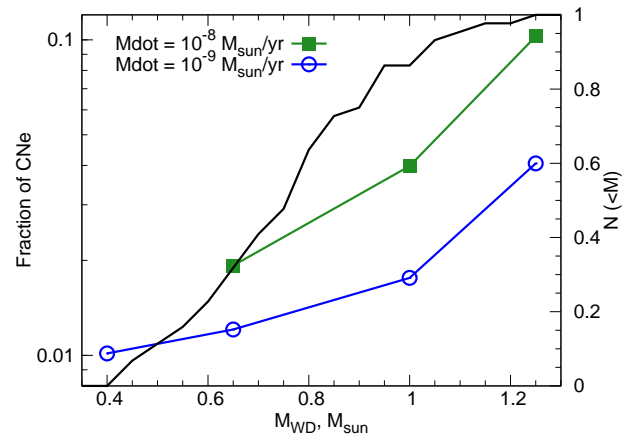}}
\caption{The upper limit of the contribution of magnetic systems to the CN frequency. The filled squares (green in the color version of this plot) correspond to an accretion rate of $ 10^{-8} \ \mathrm{M_{\odot}/yr} $, while the empty circles (blue) assumes an accretion rate of $ 10^{-9} \ \mathrm{M_{\odot}/yr} $. The solid line (black) shows the cumulative distribution of white dwarf masses in magnetic CVs (right hand \textit{y}-axis), see text for references. Note, that $ 85 $ per cent of white dwarfs are less massive than $ 1  \ \mathrm{M_{\odot}} $.} 
\label{fig:polars}
\end{figure}

\begin{table}
\caption{The list of magnetic and non-magnetic Galactic CNe, classification is taken from \citet{ritter} unless otherwise noted.}
\begin{minipage}{9cm}
\renewcommand{\arraystretch}{1}
\centering
\begin{tabular}{c c c}
\hline
Name & CV classification &  Distance, pc  \\
\hline
V603 Aql    &  SH  &  $   400^a $   \\
V1425 Aql   &  NL  &  $  2700^b $   \\
OY Ara      &  SW  &  $  1500^a $   \\
RS Car      &  SH  &  $  2100^a $   \\
V842 Cen    &  NL  &  $   500^a $   \\
RR Cha      &  SH  &  $  6100^a $   \\
V1500 Cyg   &  AM$^{\star} $  &  $  1300^a $   \\
V1974 Cyg   &  SH  &  $  1700^a $   \\
V2275 Cyg   &  IP  &  $  5500^c $    \\
V2467 Cyg   &  NL  &  $  3100^d $   \\
DM Gem      &  SW  &  $  4700^a $   \\
DQ Her      &  DQ  &  $   525^e $    \\
V446 Her    &  DN  &  $  1300^a $   \\
V533 Her    &  DQ  &  $  1300^a $    \\
DK Lac      &  VY  &  $  3900^a $   \\
BT Mon      &  SW  &  $  1000^a $   \\
GI Mon      &  IP  &  $  4600^a $    \\
GQ Mus      &  AM  &  $  5100^a $    \\
GK Per      &  DN  &  $   500^a $   \\
V Per       &  NL  &  $  1000^a $   \\
RR Pic      &  SW  &  $   500^a $   \\
CP Pup      &  SH  &  $   900^a $   \\
HZ Pup      &  NL  &  $  5800^a $   \\
V597 Pup    &  IP$^{\star} $  &  $ 10300^f $  \\
T Pyx       &  SS  &  $  2500^a $  \\
V697 Sco    &  IP  &  $ 16100^a $   \\
V373 Sct    &  IP  &  $  4600^a $   \\
WY Sge      &  DN  &  $  2000^a $  \\
V630 Sgr    &  SH  &  $   600^a $  \\
V1017 Sgr   &  DN  &  $  2600^a $  \\
V4633 Sgr   &  AM$^{\star} $&  $  9000^g $   \\
V4745 Sgr   &  IP  &  $ 14000^h $   \\ 
\hline
\end{tabular}  
\end{minipage}
$^{\star}$ The classification for V1500 Cyg, V597 Pup and V4633 Sgr was taken from \citet{stockman} \citet{warner09} and \citet{lipkin}, respectively.  \\
References are from: $^a$ \citet{shafter97} -- $^b$ \citet{kamath} -- $^c$ \citet{kiss} -- $^d$ \citet{poggiani} -- $^e$ \citet{vaytet} -- $^f$ \citet{naik} -- $^g$ \citet{lipkin} -- $^h$ \citet{csak}
\label{tab:cne}
\end{table}

\subsection{Magnetic cataclysmic variables}
\subsubsection{Upper limit on the contribution of magnetic systems to the CN rate}
\label{sec:pip}
Polars (AM Her systems) and intermediate polars (IPs) are accreting binary systems in which the accretion disk is partly (IPs) or entirely (AM Her systems) disrupted by  magnetic field of the white dwarf. These systems are sources of  relatively hard X-ray emission produced via  optically-thin bremsstrahlung in an accretion shock near the WD surface. They also may have a prominent   soft component  generated  by the  WD surface  illuminated by hard X-rays. Theory predicts \citep{lamb,king79} and observations confirm \citep{ramsay} that the soft component account for $\sim 1/2$ of the total accretion luminosity of polars. Intermediate polars may also have a soft component in their spectra of somewhat smaller luminosity, $\sim 1/3$ of the total  \citep{evans}. Correspondingly, we assume in the following calculations, that $1/2$  of the accretion  luminosity of magnetic CVs is emitted in the hard X-ray component. To the first approximation, the bremsstrahlung temperature of this component is defined by the depth of the gravitational component on the WD surface, the observed values showing considerable dispersion.  For a large sample of magnetic WDs \citep{landi,brunschweiger} we calculate   the average value of  $kT \approx 23$ keV with the standard deviation of $\approx 9$ keV.
In the following calculations we assume $kT=23$ keV and ignore the dependence of the temperature on the WD mass.
As demonstrated below   this particular choice  does not influence our results significantly.

In order to constrain the contribution of magnetic CVs to the CN frequency we compute their predicted X-ray luminosity assuming that CNe are exclusively produced in such systems.
In this calculation we assume that the presence of strong magnetic fields does not influence the characteristics of nuclear burning and results of  \cite{yaron} for the mass of the hydrogen layer $\Delta M(\dot{M}, M_{WD})$ apply. In particular, we used the $\Delta M(\dot{M}, M_{WD})$ values for the WD temperature of $T_{WD}=10^7$ K \citep{townsley}. The total accretion luminosity predicted by equation (\ref{eq:cn_lbol}) is halved (see above), corrected for interstellar absorption, and converted into the 2--10 keV energy band.   The obtained  value is then compared with the observed luminosity of the unresolved emission, corrected for the contribution of accreting WDs, as described before. The upper limit on the contribution of magnetic systems in the CN rate in M31 is shown in Fig.\ref{fig:polars}. It shows  that allowing all possible values of the WD mass, no more than $ \approx 10  $ per cent of CNe can be produced in magnetic systems. The less constraining value of the upper limit corresponds to the largest accretion rate and most massive WDs.
However, the combination of these extreme parameters is not typical for magnetic CVs as illustrated by the cumulative distribution of the WD mass in magnetic systems \citep{ritter,suleimanov,brunschweiger}. 
The distribution shows that  $\approx 85$ per cent of white dwarfs are less massive than $ 1 \ \mathrm{M_{\odot}}$. Furthermore the average accretion rate in these systems is fairly low, $ \dot{M} \sim 1.8 \cdot 10^{-9} \ \mathrm{M_{\odot}/yr} $ \citep{suleimanov}.  We conclude therefore that a more realistic upper limit should be $\approx 3$ per cent, corresponding to $ M_{WD} = 1 \ \mathrm{M_{\odot}}  $ and  $\dot{M} =  10^{-9} \ \mathrm{M_{\odot}/yr} $. 

This result weakly depends on the assumed temperature of bremsstrahlung spectrum. If  $ kT= 40 $ keV is assumed, the upper limit  becomes $\approx 4$ per cent. Only unrealisticly  large temperature of $kT=75$ keV would lead to a two times larger upper limit.

\begin{figure}
\includegraphics[width=8.75cm]{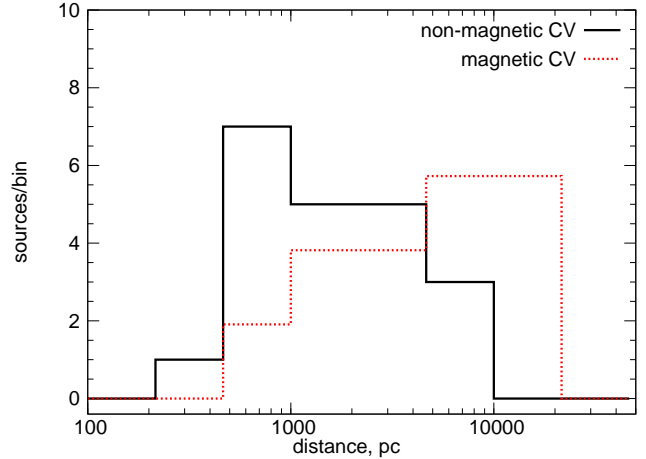}
\caption{The distance distribution for CNe arising from magnetic and non-magnetic CVs. The total number of magnetic systems is normalized by $1.9$ to match the number of non-magnetic ones. The data, used to produce this plot,  is given in Table \ref{tab:cne}.} 
\label{fig:novadistr}
\end{figure}

\subsubsection{Comparison with statistics of observed CNe from magnetic systems}

\citet{ritter}  find that $ \sim 1/3 $ of CNe, with known progenitors, are produced in CVs possessing a magnetic white dwarf (Table \ref{tab:cne}). Although this catalogue is not complete neither in the volume nor in the brightness limited sense, the $\sim 10$ times discrepancy appears to be  too large to be consistent with our upper limit of $\approx 3$ per cent.  In another study, \citet{ab} investigated the space density of magnetic and non-magnetic CVs, and  obtained that $ \sim 22  $ per cent of  all CVs are magnetic. This number  also seems to be in conflict with our upper limits. As demonstrated below, these two discrepancies are caused by the single reason -- by an order of magnitude difference in mass accretion rate between magnetic and non-magnetic system.

Indeed, the average accretion rate in magnetic CVs is $ \sim 1.8 \cdot 10^{-9} \ \mathrm{M_{\odot}/yr} $ \citep{suleimanov}, whereas for non-magnetic CNe it is $ \sim 1.3 \cdot 10^{-8} \ \mathrm{M_{\odot}/yr} $ \citep{puebla}. The obvious consequence of this difference is that in magnetic systems it takes longer to accrete the same amount of material needed to trigger the nova explosion. Moreover, at lower accretion rates $ \Delta M $ is larger by a factor of $ \sim 1.5-2 $ \citep{yaron}. The consequence of these effects is that magnetic CVs undergo CN outburst $ \sim 10-20 $ times less frequently. This  explains   the second discrepancy --  between the observed fraction of magnetic CVs and our upper limit on their contribution to the CN frequency. 

Another  consequence of the smaller accretion rate is that  CN explosions are brighter in magnetic  systems, therefore they can be observed at larger distances and are sampled from larger volume than CNe from non-magnetic progenitors. This explains the first  discrepancy -- seemingly too large fraction of CN events which progenitors are magnetic systems.  
To verify this, we selected CNe with known progenitors from the catalogue of \citet{ritter} and searched for distances in the literature. In total, we collected 32 CNe, of which 11 are from magnetic progenitors  (Table \ref{tab:cne}). Their distance distributions are shown in  Fig. \ref{fig:novadistr}.   Obviously, magnetic systems are located at larger distances: the average distance is $ \approx 6.6 $ kpc and $ \approx 2.2 $ kpc for magnetic and non-magnetic CNe. Below $ 1 $ kpc almost four times more non-magnetic CNe are observed (the normalized number of magnetic systems is $ \approx1.9 $  and $8$ for non-magnetic), whereas above $ \approx 5 $ kpc there are more magnetic CVs ($\approx11.4$ vs $3$).  The difference between two distributions is statistically significant, with the K-S probability of $\approx 0.008$.

The difference in the distance can be compared with the prediction based on the expected brightness at the maximum of the lightcurve. 
Calculations of \citet{yaron} show that for $0.65 \ \mathrm{M_\odot}$ WD  the maximum brightness achieved during the  CN explosion will differ by $\approx 1$ magnitude, the difference being smaller for a more massive WD. Taken at the face value, this would suggest the distance difference of $\sim 1.6$ which is smaller than  the observed difference in average distance by a factor of $\sim 3$. Given the crudeness of this calculation,  we conclude that these two numbers are broadly consistent with each other.

\subsection{Dwarf novae}
\label{sec:dn}
Dwarf novae (DNe) are a subclass of CVs showing frequent quasi-periodic outbursts due to thermal-viscous instability of the accretion disk \citep{osaki,hoshi}. In quiescence they become sources of relatively hard X-ray emission from the optically thin boundary layer,  whereas in outburst state the optically thick accretion disk emits predominantly in the ultraviolet and soft X-ray bands.  One of the models used to describe the emission spectra in quiescence states is  a cooling flow model \citep[e.g.][]{mukai}. Spectral analysis of a large number of quiescence spectra showed a relatively large dispersion in the value of the initial temperature $kT_{max}\sim 8-55$ keV with the average value of 23 keV \citep{pandel}.  This value is assumed in the calculations performed in this section. It is also assumed, that only half of the accretion energy is emitted in the boundary layer.

\begin{figure}
\resizebox{\hsize}{!}{\includegraphics{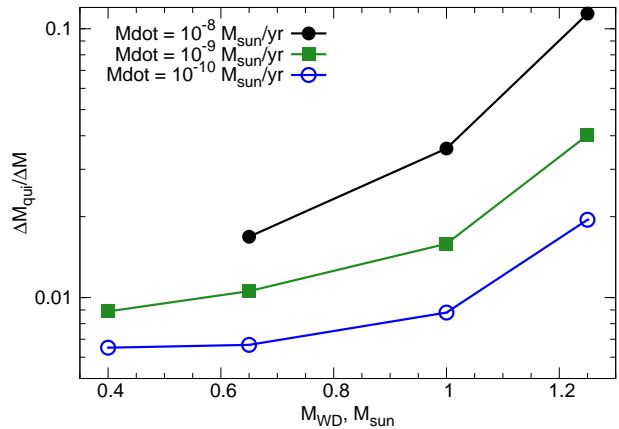}}
\caption{The fraction of mass accreted in quiescence as a function of white dwarf mass. Different mass accretion rates were assumed, the filled circles (black) correspond to $ 10^{-8} \ \mathrm{M_{\odot}/yr} $, the filled boxes (green) represent $ 10^{-9} \ \mathrm{M_{\odot}/yr} $, whereas the empty circles (blue) stand for $ 10^{-10} \ \mathrm{M_{\odot}/yr} $.} 
\label{fig:dnluminosity}
\end{figure}

Dwarf Novae give rise to CNe. On the other hand, in the quiescence state they may contribute to the unresolved emission. In the outburst state, to the opposite,  soft emission  is largely hidden because of the interstellar absorption. Therefore the bimodal spectral behavior of dwarf novae can be used to constrain the fraction of mass which is accreted in quiescence, similar to how the upper limit on the contribution of magnetic systems was obtained. Using same approach we calculate the X-ray luminosity in the quiescence state assuming that dwarf novae are responsible for a $\sim 1/2$ of CN events (the other half is due to nova-like variables, \cite{ritter}). With this value one can put an upper limit on the fraction of mass, accreted during quiescence. The result of this calculation is shown in Fig. \ref{fig:dnluminosity}. As in the case of magnetic systems, the absolute upper limit is $\sim 10$ per cent. However,  the WD mass in DNe typically does not exceed $\approx 0.9 \ \mathrm{M_\odot}$ and the mass accretion rate is not much larger that $\approx 10^{-8} \ \mathrm{M_\odot}$/yr, therefore the realistic upper limit is $\approx 3$ per cent.
As before, this result is  not strongly sensitive to the choice of $ kT_{max}$ parameter of the cooling flow model. Assuming $kT_{max} = 55 $ keV our upper limits would increase by $ \approx 1/7$ of their value.

Using this upper limit, the enhancement factor of the accretion rate in outburst state can be inferred. Obviously, this quantity depends on the fraction of time  spent in the outburst. Since for majority of DNe it is smaller than $\sim 1/5$ \citep{wils}, the accretion rate increases by  $ \sim 30 - 1000 $ during outburst. This conclusion is consistent with observations of dwarf novae. Indeed, one of the best studied dwarf nova, SS Cyg, spends $ \sim 75 $ per cent of the time in quiescence \citep{cannizzo}. The accretion rate varies from  $ 5 \cdot 10^{-11} \ \mathrm{M_{\odot}/yr} $ in quiescence \citep{urban} to  $ 3.2 \cdot 10^{-9} \ \mathrm{M_{\odot}/yr} $ in outburst \citep{hamilton}, hence the enhancement factor is $64 $ and $ \approx 95 $ per cent of the material is accreted in outburst periods, in good agreement with our results.

\subsection{Generalization of the results}
Although we considered only the bulge of M31, our results  can be generalized to other early-type galaxies. 

The X-ray to K-band luminosity ratio in early-type galaxies, galaxy bulges, and in the Milky is in good agreement in the $ 2-10 $ keV band \citep[][present work]{revnivtsev2,sazonov,m32}. 
On the other hand, for a sample of early type galaxies, including M31,  \citet{williams}. obtained fairly uniform values of the luminosity specific nova rate ($\nu_{K} $), in the range of $ (1.8-2.7) \cdot 10^{-10} \ \mathrm{L_{\odot,K} \ yr^{-1}} $. In apparent contradiction with this, we obtain $ \nu_{K} = (6.0 \pm 1.0) \cdot 10^{-10} \ \mathrm{L_{\odot,K} \ yr^{-1}} $ for the bulge of M31, based on  its CN frequency and K-band luminosity. However,  \citet{williams} computed the K-band luminosity from  (B--K) colour and B-band luminosity of the galaxy, which in case of M31 results in  by a factor of $ \sim 3 $ larger  K-band luminosity than the 2MASS value, and by a factor of $ \sim 2 $ larger than deduced from the Spitzer data \citep{williams}. With this correction, our value for the K-band specific CN rate for the bulge of M31 corresponds to  $ \nu_{K} = (3.0 \pm 0.5) \cdot 10^{-10} \ \mathrm{L_{\odot,K} \ yr^{-1}} $, which  is  in a good agreement with other galaxies. 

As  both X/K ratio and $\nu_K$ are similar, our conclusions hold for  other early type galaxies.

\section{Conclusion}
We studied unresolved X-ray emission from the bulge and disk of M31 using publicly available  \textit{XMM-Newton} and \textit{Chandra} data. The \textit{XMM-Newton} survey of M31 covered the entire galaxy with the total exposure time of $ \approx 639 \ \mathrm{ks} $ after filtering. \textit{Chandra} data covered the bulge and the southern part of the  disk, with the exposure time of the filtered data  of $ \approx 222 \ \mathrm{ks} $. 

We demonstrated that unresolved X-ray emission, associated with the bulge of the galaxy, extends into the disk with similar X-ray to K-band luminosity ratio. We obtained $ L_X/L_K = (3.4-4.5) \cdot 10^{27} \ \mathrm{erg \ s^{-1 } \ L_{\odot}^{-1}} $ in the $ 2-10 $ keV band for all studied regions in the bulge and the disk of M31, which is in good agreement  with those obtained for the Milky Way and for M32. This  suggests that the unresolved X-ray emission in M31 may have similar origin to the Galactic ridge X-ray emission, namely it is a superposition of a large number of faint compact sources, such as accreting white dwarfs and coronally active binaries.

We investigated X-ray emission associated with the 10-kpc star-forming ring based on \textit{XMM-Newton} data. We characterized this emission with $ \mathrm{L_{X}/SFR} $ ratio, where $ L_X $ is calculated for the $ 0.5-2 $ keV energy range. We found that its value is spatially variable. After excluding the two regions along the minor axis of the galaxy, which are likely contaminated by the hot gas outflow, we obtained values ranging from zero to $ \mathrm{L_{X}/SFR} = 1.8 \cdot 10^{38} \ \mathrm{(erg \ s^{-1})/(M_{\odot}/yr)} $. The origin of these variations remains unclear.

We derived constraints on the nature of Classical Nova progenitors based on the brightness of unresolved emission.  We demonstrated that magnetic CVs  --  polars and intermediate polars, do not contribute more than $ \sim 3 $ per cent to the observed CN frequency, assuming values of parameters most likely for the CN progenitors, the absolute upper limit being  $ \approx 10 $ per cent. We also showed  that in dwarf novae  $ \gtrsim 90-95 $   per cent of the material is accreted during outbursts, and only a small fraction during quiescent periods.

\bigskip
\begin{small}
 
\noindent
\textit{Acknowledgements.}
We thank the anonymous referee for useful comments.
The authors are grateful to Hans Ritter for helpful discussions of cataclysmic variables.
This research has made use of \textit{Chandra} archival data provided by 
the \textit{Chandra} X-ray Center. The publication makes use of software 
provided by the \textit{Chandra} X-ray Center (CXC) in the application 
package \begin{small}CIAO\end{small}. \textit{XMM-Newton} is an ESA science mission with instruments 
and contributions directly funded by ESA Member States and the USA (NASA).
The \textit{Spitzer Space Telescope} is operated by the Jet Propulsion Laboratory, 
California Institute of Technology, under contract with the 
National Aeronautics and Space Administration. 
\end{small}

\end{document}